\documentclass[runningheads]{llncs}
\usepackage{graphicx}

\usepackage{amsmath}
\usepackage{amssymb}
\usepackage{booktabs}
\usepackage{tabularx}
\usepackage{multirow}
\usepackage{arydshln}
\usepackage{todonotes}
\usepackage{subcaption}
\usepackage{soul}
\soulregister\cite7
\soulregister\ref7

\newcommand\blfootnote[1]{%
  \begingroup
  \renewcommand\thefootnote{}\footnote{#1}%
  \endgroup
}

\begin{document}


\title{A persistent homology-based topological \\ loss function for multi-class CNN \\ segmentation of cardiac MRI \thanks{Nick Byrne is funded by a National Institute for Health Research (NIHR), Doctoral Research Fellowship for this research project. This report presents independent research funded by the NIHR. The views expressed are those of the author(s) and not necessarily those of the NHS, the NIHR or the Department of Health and Social Care. The authors have no conflicts of interest to disclose.}}
\titlerunning{A topological loss for multi-class CMR segmentation}

\author{Nick Byrne\inst{1,2} \and
James R. Clough\inst{2} \and
Giovanni Montana\inst{3}$^{\dagger}$ \and
Andrew P. King\inst{2}$^{\dagger}$
}

\authorrunning{N. Byrne et al.}

\institute{Medical Physics, Guy's and St. Thomas' NHS Foundation Trust, London, UK \and
School of Biomedical Engineering \& Imaging Sciences, King's College London, UK \and
Warwick Manufacturing Group, University of Warwick, Coventry, UK
\email{nicholas.byrne@kcl.ac.uk}}

\maketitle


\begin{abstract}
With respect to spatial overlap, CNN-based segmentation of short axis cardiovascular magnetic resonance (CMR) images has achieved a level of performance consistent with inter observer variation.
However, conventional training procedures frequently depend on pixel-wise loss functions, limiting optimisation with respect to extended or global features.
As a result, inferred segmentations can lack spatial coherence, including spurious connected components or holes.
Such results are implausible, violating the anticipated topology of image segments, which is frequently known a priori.
Addressing this challenge, published work has employed persistent homology, constructing topological loss functions for the evaluation of image segments against an explicit prior.
Building a richer description of segmentation topology by considering all possible labels and label pairs, we extend these losses to the task of multi-class segmentation.
These topological priors allow us to resolve all topological errors in a subset of 150 examples from the ACDC short axis CMR training data set, without sacrificing overlap performance.

\end{abstract}

\blfootnote{\hspace{-2.5mm}$^{\dagger}$ \emph{Joint last authors.}}
\addtocounter{footnote}{-1}
\addtocounter{footnote}{-1}
\addtocounter{footnote}{-1}
\addtocounter{footnote}{-1}


\section{Introduction}

Medical image segmentation is a prerequisite of many pipelines dedicated to the analysis of clinical data.
In cardiac magnetic resonance (CMR) imaging, segmentation of short axis cine images into myocardial, and left and right ventricular components permits quantitative assessment of important clinical indices derived from ventricular volumes \cite{Ruijsink2019}.
Motivated by the burden of manual segmentation, researchers have sought automated solutions to this task.

To this end, deep learning, and in particular convolutional neural networks (CNNs), have brought significant progress \cite{Chen2020}.
One key to their success has been the design of specialised architectures dedicated to image segmentation.
Theoretically, these permit the learning of image features related to extended spatial context, such as anatomical morphology.

Whilst considerable effort has gone into investigating methods for the extraction of multi-scale image features, less attention has been paid to their role in network optimisation \cite{Duan2019}.
Instead, CNNs have often been trained using pixel-wise loss functions based on cross-entropy (CE) or the Dice Similarity Coefficient (DSC).
Though having favourable numerical properties, these are insensitive to higher order image features such as topology.
In the absence of this information, CNN optimisation can result in predicted segmentations with unrealistic properties such as spurious connected components or holes \cite{Painchaud2020}.
These errors can appear nonsensical, violating the most basic features of image segments.
If small, the presence of such errors may not preclude a high spatial overlap with the ground truth and have little consequence for certain clinical indices.
However, a wider array of downstream applications such as biophysical modelling or 3D printing, demand a high fidelity representation of such features.

In short axis CMR, prior knowledge dictates that the right ventricular cavity is bound to the left ventricular myocardium, which in turn fully surrounds the left ventricular blood pool.
In contrast to pixel-wise objectives, this is a global description of segmentation coherence.
However, whilst these constraints are simple to express qualitatively, the opaque nature of CNNs has made it difficult to explicitly exploit such prior information in model optimisation.


\subsection{Related work}
\label{subsec:previous_work}
At least in the context of large, homogeneous training datasets, CNN-based short axis segmentation has achieved a level of performance consistent with inter observer variation \cite{Chen2020}.
However, in studies of cardiovascular disease (for which datasets typically contain fewer subjects who exhibit greater morphological variation), a deficit remains.
This gap is in part characterised by the anatomically implausible errors described.

To address these modes of failure, previous works have sought to leverage prior information.
The combination of deep learning with atlas-based segmentation \cite{Duan2019} and active contour refinement \cite{Avendi2017} have both been investigated.
Whilst these extensions improve performance, their capacity to represent pathological variation in image features is limited.
Accordingly, others have injected prior information directly into CNN optimisation, developing a supervisory signal from a learned, latent distribution of plausible anatomy \cite{Oktay2018}.
Their implicit embedding, however, hinders an understanding of the extent to which such priors are related to morphology or topology as claimed.
Bridging this gap, Painchaud et al. augmented the latent space via a rejection sampling procedure, maintaining only those cases satisfying sixteen criteria related to anatomical plausibility \cite{Painchaud2020}.

Eleven of Painchaud et al.'s criteria concern anticipated anatomical topology.
Structured losses have previously been designed to capture aspects of segmentation topology including hierarchical class containment \cite{BenTaieb2016} and adjacency \cite{Ganaye2018,He2019}.
More recently, CNN-based segmentation has benefited from the global, exhaustive and robust description of topology provided by persistent homology (PH).
PH admits the construction of topological loss functions which, in contrast to those built on a latent representation of anatomical shape, allow evaluation against an explicit topological prior.
Applications have included segmenting the tree-like topology of the murine neurovasculature \cite{Haft-Javaherian2020}; and the toroidal topology of the myocardium in short axis CMR \cite{Clough2019a,Clough2019}.

\subsection{Contributions}
\label{subsec:contributions}
To the best of our knowledge, no PH-based loss function has been proposed for the task of multi-class segmentation.
Compared with the binary case, extension to this setting considers a richer set of topological priors, including hierarchical class containment and adjacency.
short axis CMR segmentation is a useful test bed for this task, not only for its clinical significance, but also for its economic representation of such priors in a well-defined, four-class problem.
In this context, our contributions are as follows:
\begin{enumerate}
  
  \item We propose a novel topological loss function, based on PH, for the task of multi-class image segmentation.
  
  \item We use the novel loss function for CNN-based segmentation of short axis CMR images from the ACDC dataset.

  \item We demonstrate significant improvement in segmentation topology without degradation in spatial overlap performance.

\end{enumerate}


\section{Materials and methods}

We address a multi-class segmentation task, seeking a meaningful division of the 2D short axis CMR image $\mathbf{X}: \mathbb{R} \times \mathbb{R} \rightarrow \mathbb{R}$ into background, right and left ventricular cavities and left ventricular myocardium\footnote{The semantic classes of this task match those of the ACDC image segmentation challenge \cite{Bernard2018}. Whilst neither considers the right ventricular myocardium, this could easily be incorporated into the framework set out in Sections \ref{subsec:multi_class_priors} and \ref{subsec:topo_loss_function}.}.
We denote the ground truth image segmentation by $\mathbf{Y}: \mathbb{R}\times \mathbb{R}\rightarrow \{ 0, 1 \} ^4$, being made up by four mutually exclusive class label maps: $Y_{bg}$, $Y_{rv}$, $Y_{my}$ and $Y_{lv}$.
We consider a deep learning solution, optimising the parameters, $\theta$, of a CNN to infer the probabilistic segmentation, $\mathbf{\hat{Y}}: \mathbb{R}\times \mathbb{R}\rightarrow [ 0, 1 ] ^4$, a distribution over the per class segmentation maps: $\hat{Y}_{bg}$, $\hat{Y}_{rv}$, $\hat{Y}_{my}$ and $\hat{Y}_{lv}$.
We write segmentation inference as $\mathbf{\hat{Y}}=f(\mathbf{X};\mathbf{\theta})$.
Given the success of CNN-based solutions, we assume that, at least with respect to spatial overlap, $\mathbf{\hat{Y}}$ is a reasonable estimate of $\mathbf{Y}$.
In this setting we describe our CNN post-processing framework for the correction of inferred segmentation topology.

\subsection{Multi-class topological priors}
\label{subsec:multi_class_priors}
In 2D, objects with differing topology can be distinguished by the first two Betti numbers: $\mathbf{b} = (b_0, b_1)$.
Intuitively, $b_0$ counts the number of connected components which make up an object, and $b_1$, the number of holes contained \cite{Otter2017}.
The Betti numbers are topological invariants permitting the specification of priors for the description of foreground image segments.
Consider our short axis example:

\begin{equation*}
\begin{aligned}[c]
\mathbf{b}^{rv} = (1, 0) \\
\mathbf{b}^{my} = (1, 1) && \qquad \text{(1)} \stepcounter{equation}\\
\mathbf{b}^{lv} = (1, 0) \\
\end{aligned}
\qquad\qquad\qquad
\begin{aligned}[c]
\mathbf{b}^{rv \cup my} = (1, 1) \\
\mathbf{b}^{rv \cup lv} = (2, 0) && \qquad \text{(2)} \stepcounter{equation}\\
\mathbf{b}^{my \cup lv} = (1, 0) \\
\end{aligned}
\end{equation*}

Equation set (1) specifies that each of the right ventricle, myocardium and left ventricle should comprise a single connected component, and that the myocardium should contain a single hole.
However, these equations only provide a topological specification in a segment-wise, binary fashion: they fail to capture inter-class topological relationships between cardiovascular anatomy.
For instance, they make no specification that the myocardium surround the left ventricular cavity or that the right ventricle and myocardium should be adjacent.

By the inclusion-exclusion principle, the topology of a multi-class image segmentation is characterised by that of all foreground objects and all possible object pairs: see Equation set (2).
For convenience, we collect Equation sets (1) and (2) into a 3D Betti array $\mathbf{B}: \{1, 2, 3\} \times \{1, 2, 3\} \times \{0, 1\} \rightarrow \mathbb{R}$.
Each element $B^{ij}_{d}$\footnote{In $B^{ij}_{d}$ we divide indices between sub and super scripts to make clear the difference between class labels ($i$, $j$) and topological dimension ($d$), without further significance.} denotes the Betti number of dimension $d$ for the ground truth segmentation $Y_{i \cup j}$\footnote{We use the union operator ($\cup$) to combine individual classes of a multi-class segmentation.
When applied to a binary segmentation, $Y_{i \cup j}$ is the pixel-wise Boolean union of classes $i$ and $j$.
When applied to a probabilistic segmentation, $\hat{Y}_{i \cup j}$ is the pixel-wise probability of class $i$ or $j$.
We consider the union of a class with itself to be the segmentation of the single class: $Y_{i \cup j=i} = Y_{i}$ and $\hat{Y}_{i \cup j=i} = \hat{Y}_{i}$.}.
Vitally, even in the absence of the ground truth, $\mathbf{B}$ can be determined by prior knowledge of the anatomy to be segmented.

\subsection{Topological loss function}
\label{subsec:topo_loss_function}
To expose topological features we apply PH (see \cite{Otter2017} for a theoretical background).
For a practical understanding of the topological loss described, the results of PH analyses are most easily appreciated by inspection of persistence barcodes.
The PH barcode summarises the topological features present within data.
However, rather than providing a singular topological description, the barcode returns a dynamic characterisation of the way that the topology evolves as a function of some scale parameter.
More concretely, and in our context a barcode reflects the topology of a probabilistic segmentation $\hat{S}: \mathbb{R}\times \mathbb{R}\rightarrow [ 0, 1 ]$, binarised at all possible probability thresholds in the range $[0, 1]$.
As the threshold, $p$, reduces, the barcode diagram tracks the evolving topology of the binarised segmentation $\hat{S}_{p}$.

\begin{figure}[t]
\centering
\includegraphics[width=1.0\linewidth]{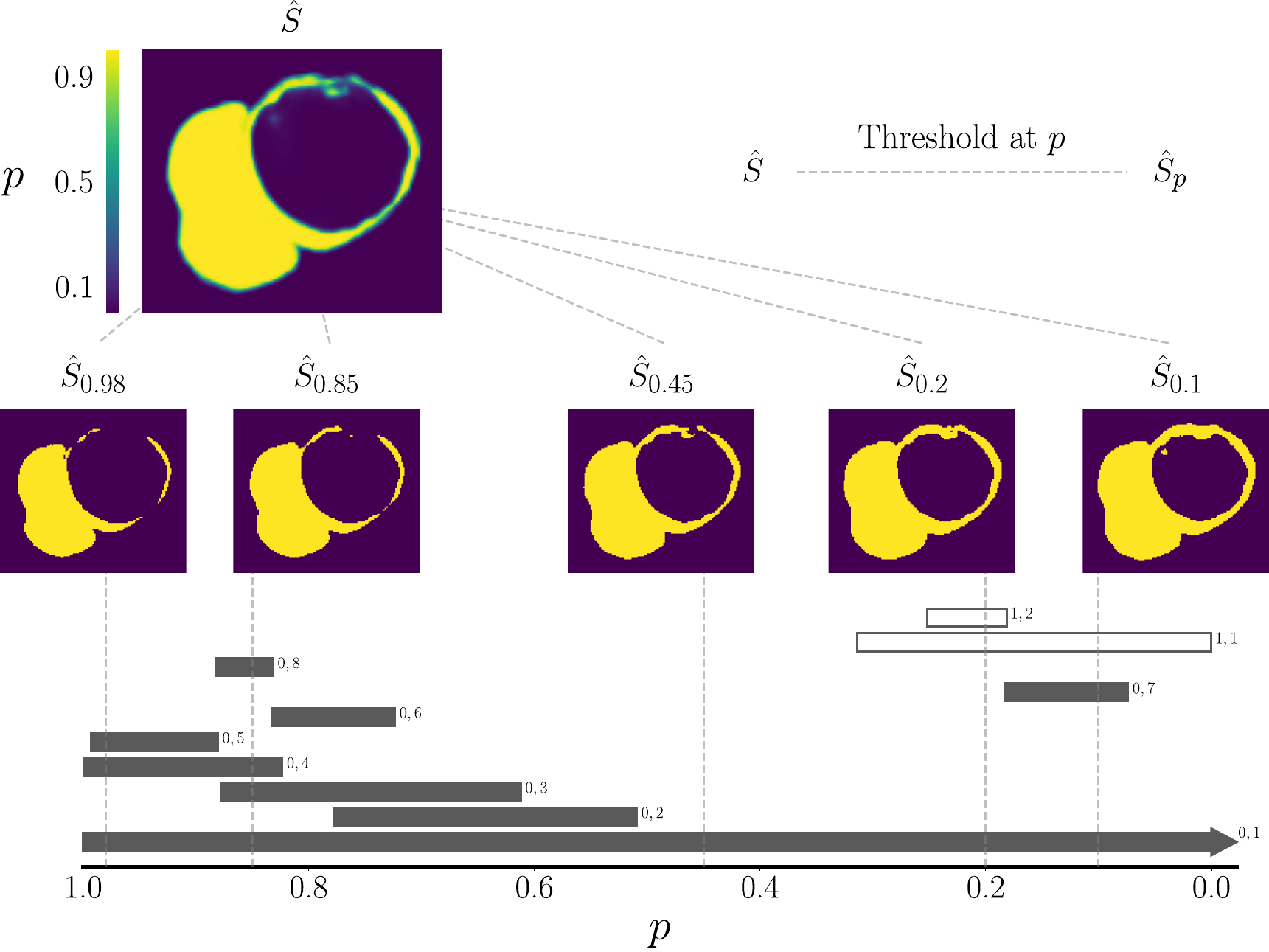}
\caption{Construction of the PH barcode. The barcode reflects the topological features of the probabilistic segmentation $\hat{S}$, when binarised at all possible probability thresholds in the interval $[0, 1]$. At a particular $p$: the number of vertically intersected solid bars counts connected components ($d = 0$); open bars count the number of loops ($d = 1$). Additionally, each bar is labelled with its topological dimension, and its persistence ranking in order of descending lifetime: $d, l$.} 
\label{fig:PH}
\end{figure}

\begin{figure}[p]
\centering
\includegraphics[width=1.0\linewidth]{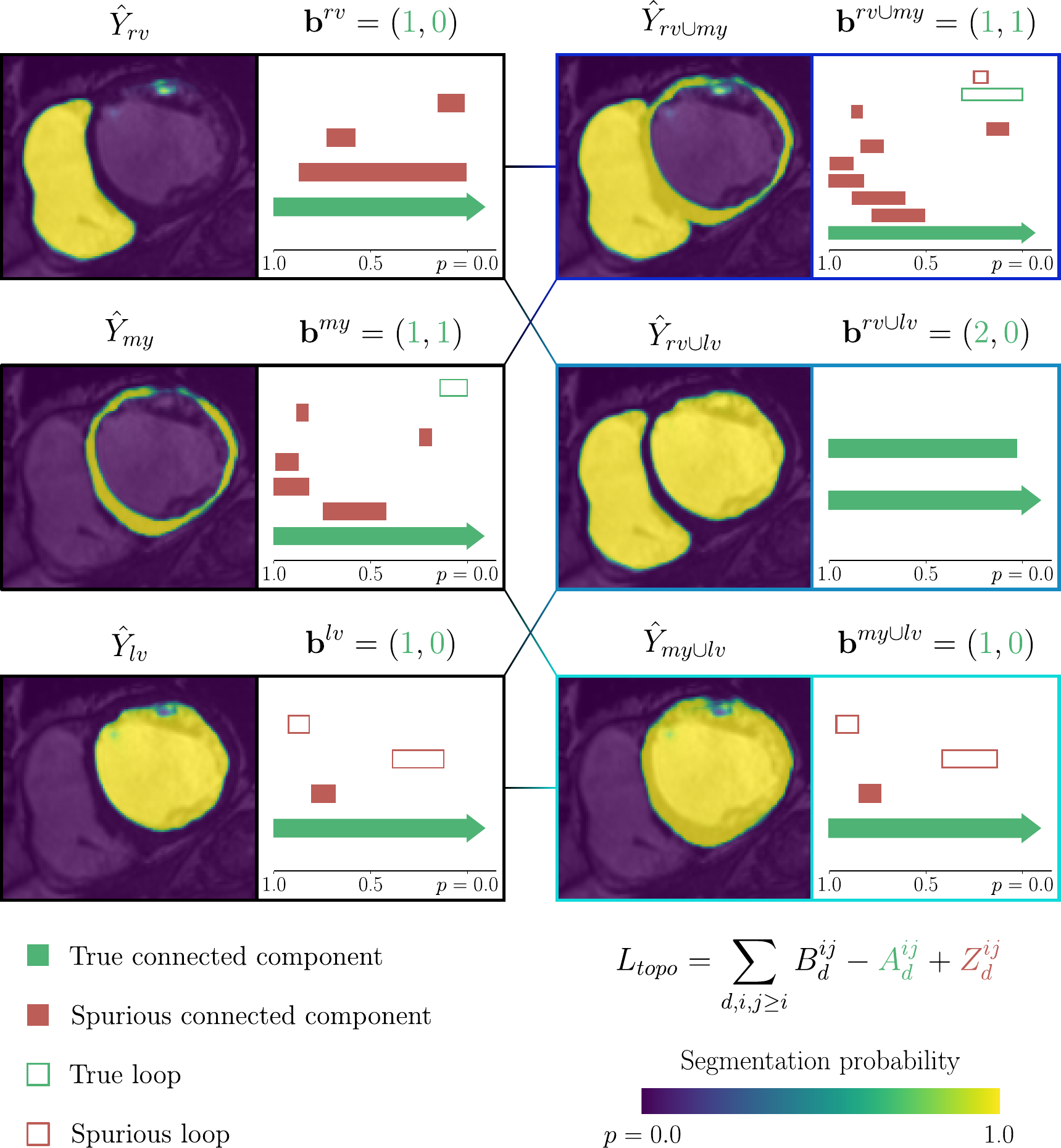}
\caption{Construction of the loss $L_{topo}$. Each probabilistic segmentation ($Y_{i}$ or $Y_{i \cup j}$), is accompanied by its associated persistence barcode (for clarity, only features with a lifetime $\Delta p_{d,l} \ge 0.05$ are shown).
$L_{topo}$ weighs the persistence of topological features which match the topological description ($A^{ij}_{d}$; depicted as green bars), against those which do not ($Z^{ij}_{d}$; depicted as red bars).
To sensitise $L_{topo}$ to multi-class label map topology, the result is repeated for, and summed over all topological dimensions ($d$), and individual and paired label sets ($i,j \ge i$).}
\label{fig:L_topo}
\end{figure}

Critical values of $p$ admit changes in the topological features of $\hat{S}_{p}$.
In Figure \ref{fig:PH}, such values are indicated by the endpoints of each bar.
Accordingly, the persistence of a topological feature $\Delta p$ is the length of its associated bar.
Moreover, the presentation of each bar indicates the topological dimension of the feature shown: solid bars are connected components; open bars are loops.
Persistent bars are considered robust to small perturbations, suggesting that they are true topological features of the data.
Hence, in Figures \ref{fig:PH} and \ref{fig:L_topo}, we consider barcodes in order of descending lifetime after grouping by topological dimension.
From the persistence barcode of the probabilistic segmentation $\hat{S}$, we write the lifetime of the $l^{th}$ most persistent feature of dimension $d$ as $\Delta p_{d,l}(\hat{S})$.

Importantly, topological persistence can be determined in a fashion that is differentiable and consistent with gradient-based learning \cite{Bruel-Gabrielsson2019}.
This permits the construction of topological loss functions, exposing the differences between $\hat{\mathbf{Y}}$ and our prior specification $\mathbf{B}$.
Key to our formulation is the choice of the probabilistic segmentation $S$, from which we extract topological features.
To align with the theory of Section \ref{subsec:multi_class_priors}, we consider the persistence barcode for all foreground class labels and class label pairs (see Figure \ref{fig:L_topo}):
\vspace{-2mm}
\begin{equation} \label{eq:L_topo}
\begin{aligned}[c]
\\
L_{topo}  = \sum_{d, i, j \geq i} B_{d}^{ij} - A^{ij}_{d} + Z^{ij}_{d} \\
\\
\end{aligned}
\qquad\qquad
\begin{aligned}[c]
A^{ij}_{d} \;\; &= \;\;\; \sum_{l=1}^{B_{d}^{ij}} \;\;\;\;\! \Delta p_{d,l}(\hat{Y}_{i \cup j}) \\
Z^{ij}_{d} \;\; &= \sum_{l=B_{d}^{ij} + 1}^{\infty} \Delta p_{d,l}(\hat{Y}_{i \cup j})\\
\end{aligned}
\end{equation}

$A^{ij}_{d}$ evaluates the total persistence of the $B_{d}^{ij}$ longest, $d$-dimensional bars for the probabilistic union of segmentations for classes $i$ and $j$, $\hat{Y}_{i \cup j}$ (see Footnote 2).
Assuming that the inferred segmentation closely approximates the ground truth, and recalling that $l$ ranks topological features in descending order of persistence, $A^{ij}_{d}$ measures the presence of anatomically meaningful topological features.
$Z^{ij}_{d}$ evaluates the persistence of spurious topological features that are superfluous to $B_{d}^{ij}$.
Summing over all topological dimensions $d$, and considering all class labels $i,j=i$ and class label pairs $i,j>i$, optimising $L_{topo}$ maximises the persistence of topological features which match the prior specification, and minimises those which do not.

As in the single class formulation presented in \cite{Clough2019}, $L_{topo}$ is used to guide test time adaptation of the weights of a pre-trained CNN $f(\mathbf{X}; \theta)$, seeking an improvement in inferred segmentation topology.
A new set of network parameters $\theta_{n}$ are learned for the individual test case $\mathbf{X}_{n}$.
However, since topology is a global property, there are many segmentations that potentially minimise $L_{topo}$.
Hence, where $V_{n}$ is the number of pixels in $\mathbf{X}_{n}$, a similarity constraint limits test time adaptation to the minimal set of modifications necessary to align the segmentation and the topological prior, $\mathbf{B}$:

\begin{equation} \label{eq:ltp}
L_{TP} = L_{topo}(f(\mathbf{X}_{n}; \theta_{n}), \mathbf{B}) + \frac{\lambda}{V_{n}} |f(\mathbf{X}_{n}; \theta) - f(\mathbf{X}_{n}; \theta_{n})|^{2}
\end{equation}

\subsection{Implementation}
\label{subsec:implementation}
We apply our loss to a topologically consistent subset of the ACDC \cite{Bernard2018} training data.
Ignoring irregular anatomical appearances at apex and base, we extract three mid ventricular slices from each short axis stack, including end diastolic and systolic frames.
We achieve a dataset of 600 examples from 100 subjects.
As per the winning submission to the ACDC Challenge, all image - label pairs were resampled to an isotropic pixel spacing of 1.25~mm (less than the mean and median spatial resolution of the training data) and normalised to have zero mean and unit variance \cite{Isensee2018}.
Subjects were randomly divided between training, validation and test sets in the ratio 2:1:1, stratified by diagnostic group according to ACDC classification.

A U-Net model \cite{Ronneberger2015} was trained using CE loss and the combined training and validation set of 450 examples, for 16,000 iterations.
We used the Adam optimiser with a learning rate of $10^{-3}$.
Each minibatch contained ten patches of size 352~by~352, randomly cropped from ten different patients, zero padding where necessary.
Data augmentation applied random rotations between $\pm 15^{\circ}$.
Graphics processing unit (GPU)-accelerated training took nine hours.

Topological post-processing was performed on the inferred multi-class segmentations of the held-out test set.
This sought to minimise $L_{TP}$ for the topological priors expressed by Equation sets (1) and (2), and summarised in $\mathbf{B}$.
In Equation \eqref{eq:ltp} we used a value of $\lambda = 1000$.
Test time adaptation used the Adam optimiser with a learning rate of $10^{-5}$ for 100 iterations.
In the worst case, topological post-processing required six minutes per short axis slice.

All experiments were implemented using PyTorch: making use of the Topolayer package introduced in \cite{Bruel-Gabrielsson2019} for the computation of topological persistence.
The reported hyperparameters for both supervised training and topological post-processing were optimised using the validation set of 150 examples.


\section{Results and Discussion}

We assess multi-class topological post-processing against several baselines.
Each applies a variant of connected component analysis ($CCA$) or topological post-processing ($TP$) to the segmentation inferred by the U-Net trained in a fully supervised fashion ($U\!Net$).
In all cases, a discrete segmentation is finally achieved by the set of labels which maximise inferred probability on a pixel-wise basis:

\noindent
\begin{center}
\begin{tabular}{ l l }
  $U\!Net$~                      & ~the output of the U-Net trained in a supervised manner. \\
  $U\!Net+ CCA$~                 & ~the largest connected components for each foreground label. \\
  $U\!Net + {TP}_{i, j = i}$~    & ~our topological loss based on individual class labels only. \\
  $U\!Net + {TP}_{i, j \geq i}$~ & ~our topological loss based on individual and paired labels.
\end{tabular}
\end{center}

Table \ref{tab:results} presents results on the held-out test set.
Spatial overlap is quantified by DSC, averaged across cardiac phases.
Topological performance is assessed as the proportion of cases in which the discrete segmentation demonstrated the correct multi-class topology: inferred and ground truth segmentations shared the same set of Betti numbers for individual and paired labels.
Finally, we summarise the effect of any post-processing on spatial overlap by the change in DSC.

\newcolumntype{C}{>{\centering\arraybackslash}X}
\newcolumntype{L}{>{\raggedright\arraybackslash}X}
\newcolumntype{R}{>{\raggedleft\arraybackslash}X}
\begin{table}[t]
\centering
\caption{Segmentation results on held-out test set. Spatial overlap: Dice Similarity Coefficient ($DSC$) per class; the average over classes ($\mu$); and the change induced by post-processing ($\Delta\mu$). Topological accuracy: $T$ is the proportion of test images with the correct multi-class topology. $\sigma$ is the standard deviation.}
\label{tab:results}
\begin{tabularx}{\textwidth}{ *{6}{L} c }
\toprule
& \multicolumn{5}{c}{$DSC_{(\sigma)}$} & \multirow{2}{*}{$T_{(\sigma)}$} \\
\cmidrule{2-6}
& $rv$ & $my$ & $lv$ & $\mu$ & $\Delta\mu$ &   \\
\midrule
\multicolumn{1}{l}{$U\!Net$} & $0.891_{(0.115)}$ & $0.885_{(0.050)}$ & $0.954_{(0.039)}$ & $0.910_{(0.045)}$ & ---               & $0.853_{(0.032)}$ \\
$+ CCA$                      & $0.892_{(0.113)}$ & $0.886_{(0.049)}$ & $0.954_{(0.039)}$ & $0.911_{(0.044)}$ & $0.001_{(0.004)}$ & $0.927_{(0.024)}$ \\
$+ {TP}_{i, j = i}$          & $0.889_{(0.130)}$ & $0.887_{(0.048)}$ & $0.954_{(0.039)}$ & $0.910_{(0.049)}$ & $0.000_{(0.012)}$ & $0.980_{(0.013)}$ \\
$+ {TP}_{i, j \geq i}$       & $0.889_{(0.129)}$ & $0.888_{(0.048)}$ & $0.954_{(0.039)}$ & $0.910_{(0.049)}$ & $0.000_{(0.011)}$ & $1.000_{(0.000)}$ \\
\bottomrule
\end{tabularx}
\end{table}

\begin{figure}[b]
\centering
\begin{subfigure}{0.3\textwidth}
  \centering
  \includegraphics[width=\linewidth]{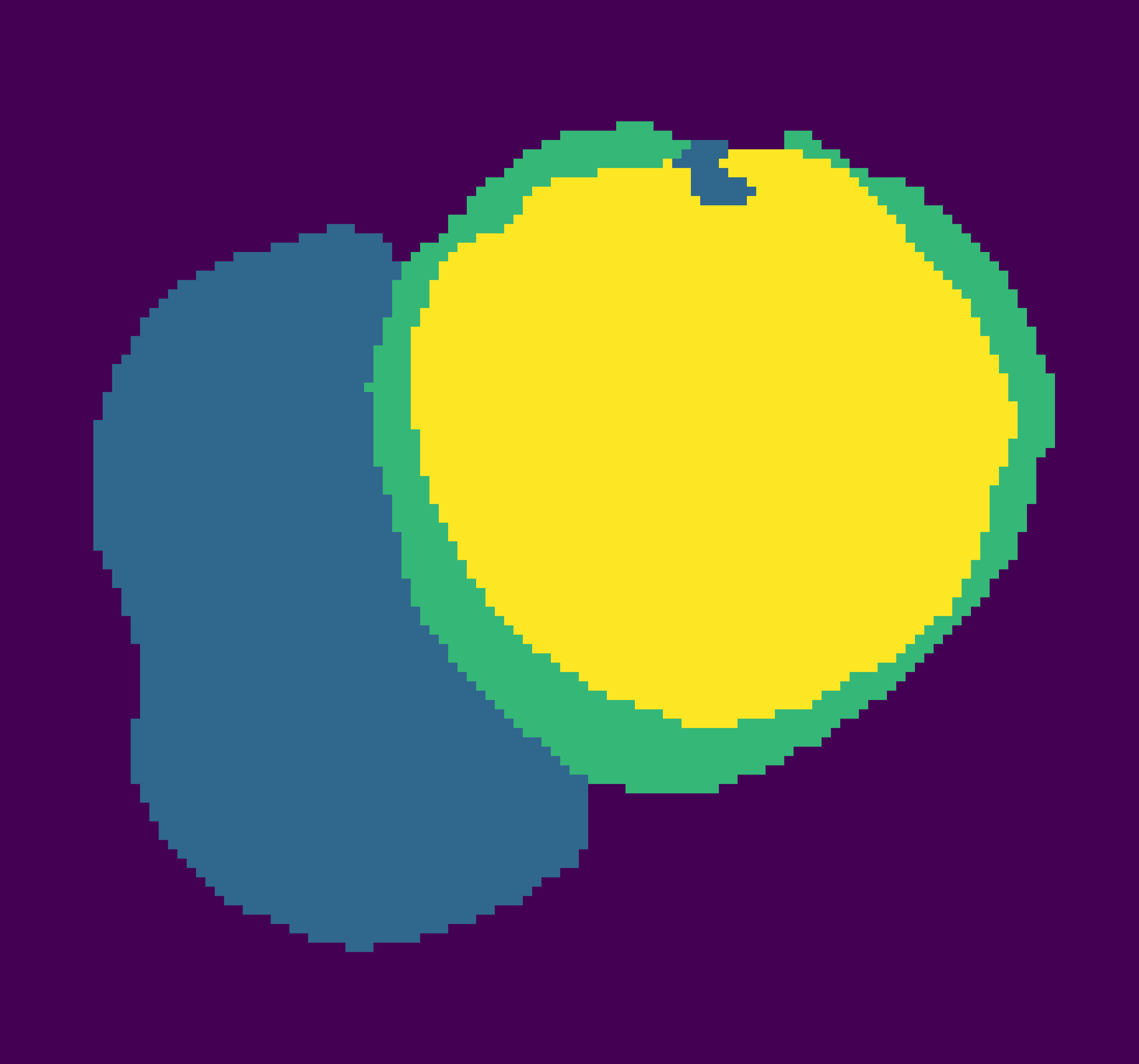}
  \caption{$UNet$}
\end{subfigure}
\begin{subfigure}{0.3\textwidth}
  \centering
  \includegraphics[width=\linewidth]{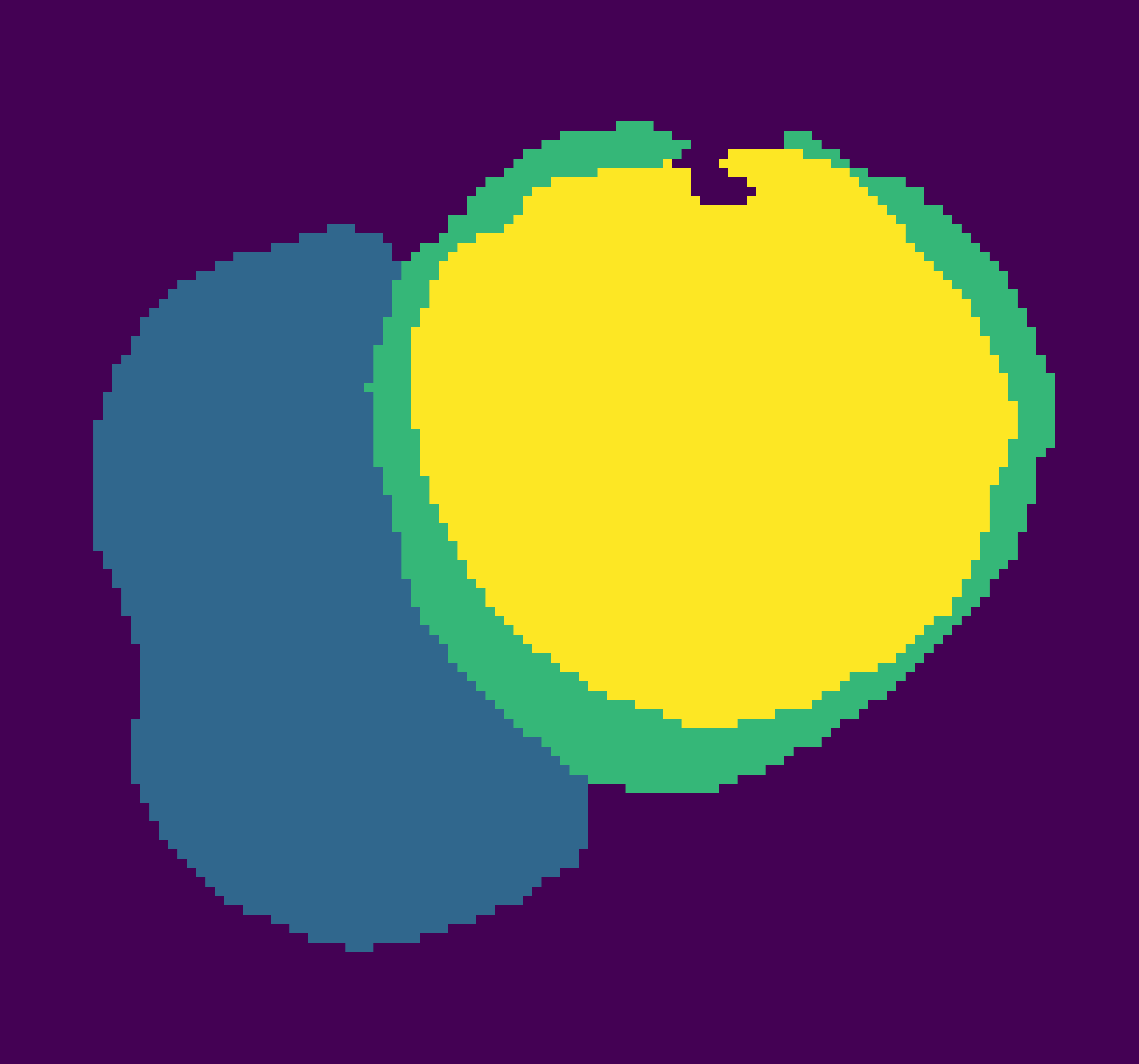}
  \caption{$CCA$}
\end{subfigure}
\begin{subfigure}{0.3\textwidth}
  \centering
  \includegraphics[width=\linewidth]{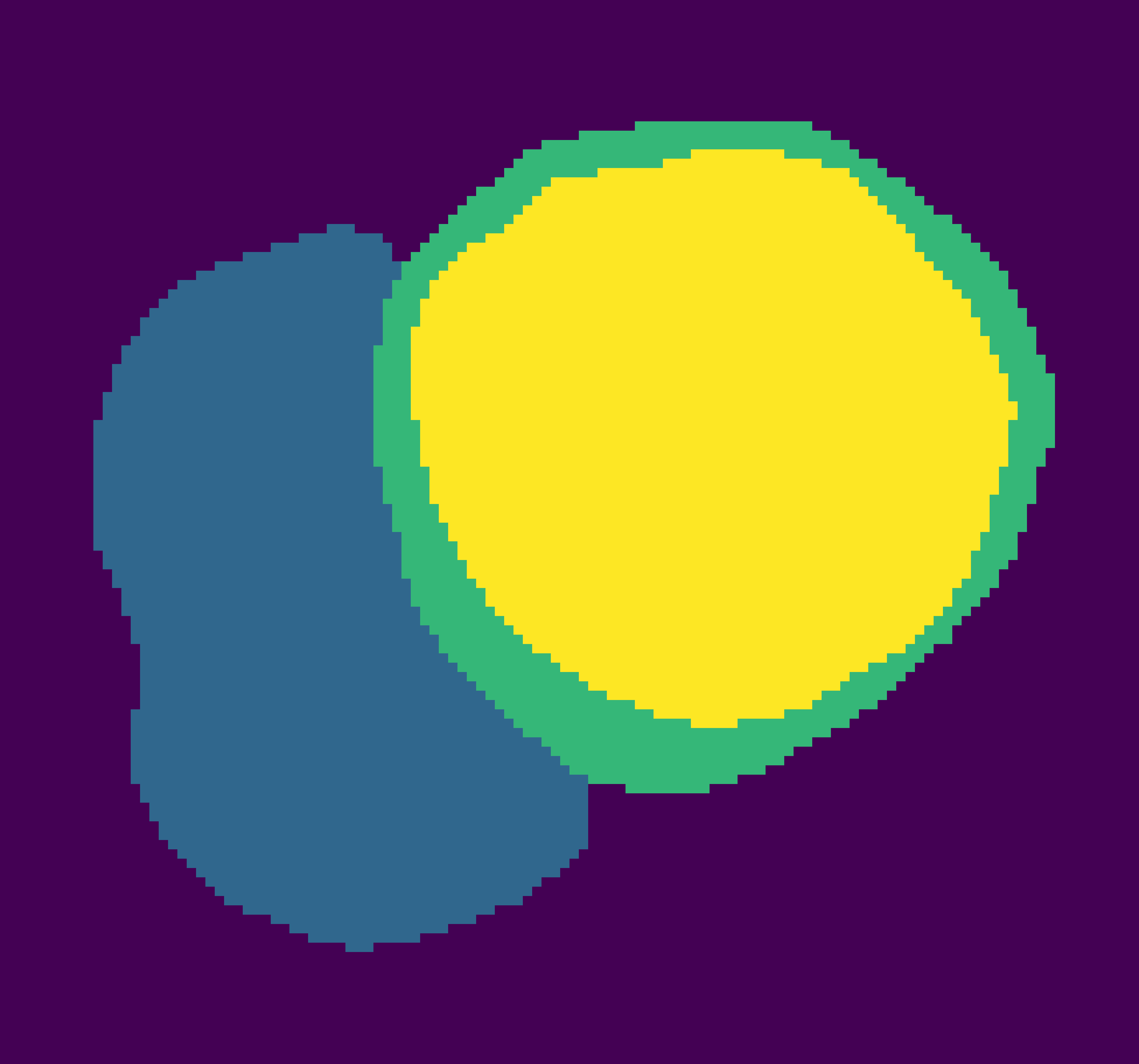}
  \caption{$TP_{i, j \geq i}$}
\end{subfigure}
\caption{Topological post-processing enables expressive correction of U-Net errors.}
\label{fig:expression}
\end{figure}

\begin{figure}[t]
\centering
\begin{subfigure}{0.3\textwidth}
  \centering
  \includegraphics[width=\linewidth]{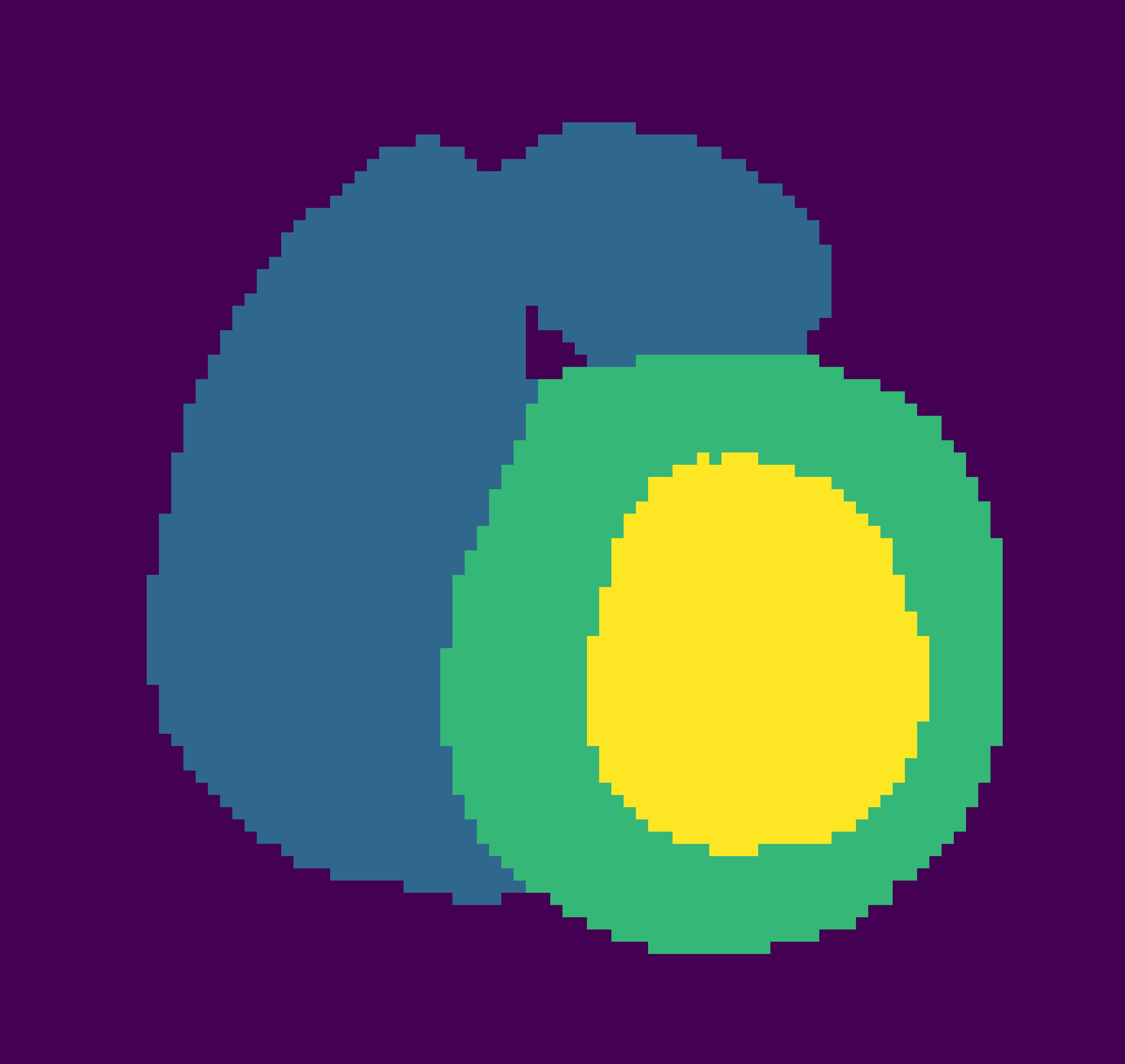}
  \caption{$UNet$}
\end{subfigure}
\begin{subfigure}{0.3\textwidth}
  \centering
  \includegraphics[width=\linewidth]{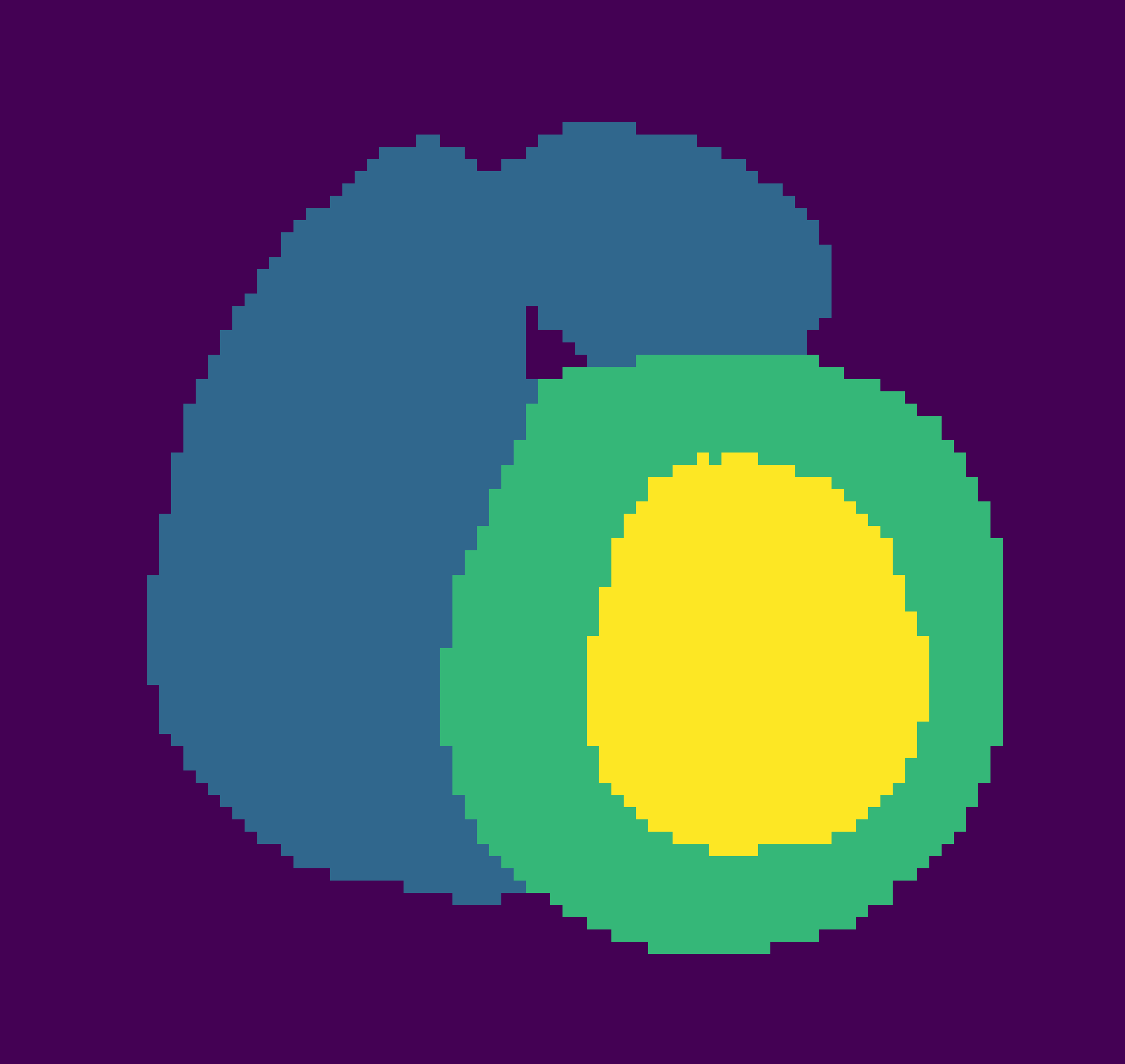}
  \caption{$+ {TP}_{i, j = i}$}
\end{subfigure}
\begin{subfigure}{0.3\textwidth}
  \centering
  \includegraphics[width=\linewidth]{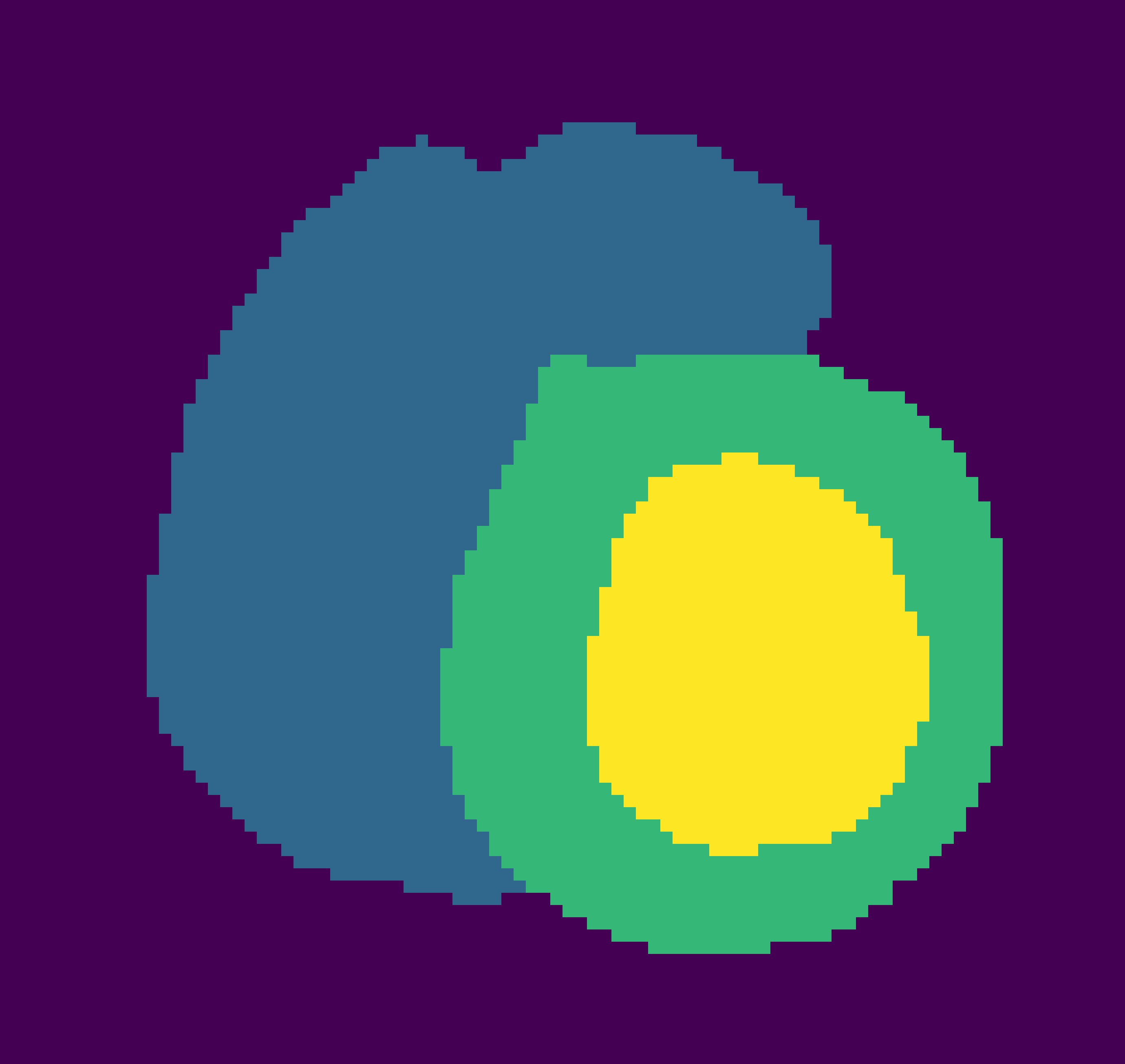}
  \caption{$+ {TP}_{i, j \geq i}$}
\end{subfigure}
\caption{Multi-class topological priors capture a rich topological description.}
\label{fig:incremental}
\end{figure}

Table \ref{tab:results} confirms that pixel-wise metrics of spatial overlap do not reliably predict topological performance.
Trained by CE, the supervised model infers a segmentation with incorrect topology in almost $15\%$ of cases.
Whilst spurious connected components alone account for approximately $50\%$ of U-Net errors, higher dimensional topological errors are not insignificant.
As commonly employed, CCA is insensitive to such features, resolving spurious components by their discrete removal.
Our method takes a probabilistic approach to the treatment of priors, modifying inferred segmentations by CNN parameter update.
This permits expressive topological refinement, as illustrated in Figure \ref{fig:expression}.

Optimisation with respect to topological priors for individual labels resolves half of high dimensional errors.
This approach, $+ {TP}_{i, j = i}$, reflects the naive extension of the binary segmentation method outlined in \cite{Clough2019}.
However, by exact binomial test, and after Bonferroni correction, there was no statistically significant difference (95\% confidence) in the proportion of topological errors between $+ {TP}_{i, j = i}$ and $+ CCA$ ($p=0.025\times6$).
The best performing scheme is our proposed model, $+ {TP}_{i, j \geq i}$, which considers priors for all individual and paired labels.
The incremental benefit of $+ {TP}_{i, j \geq i}$ is shown in Figure \ref{fig:incremental}.
Significantly, and without degradation in DSC, our approach resolves all topological errors which remain after CCA ($p=0.001\times6$).

Compared with losses based on a learned latent space of plausible anatomical shapes \cite{Oktay2018,Painchaud2020}, PH-based loss functions allow optimisation with respect to an explicit topological prior.
This is beneficial in terms of interpretability and in low data settings.
However, PH-based losses also allow topological prior information to be decoupled from its appearance in training data, on which a learned distribution is necessarily biased.
We speculate that additional biases may degrade performance when used to refine the segmentation of out-of-sample test data.
Favourably, PH-based priors permit even-handed topological post-processing in the presence of pathology-induced structural variation.
At the same time, we recognise the potential complementarity of these approaches if explicit topological specification could be enhanced with learned shape priors.

More generally, the PH losses described are limited by their reliance on an expert-provided prior.
Demonstrating our approach on mid-ventricular slices permitted the consistent application of Equation sets (1) and (2).
Allowing for a slice-wise specification, the same approach could equally be applied to apical and basal slices, including associated topological changes.
At present this would necessitate operator intervention.
However, we also observe that this requirement is an artefact of CMR short axis acquisition.
Given 3D cine data, we could specify priors which truly reflect the topology of anatomy rather than its appearance within a 2D acquisition.


\section{Conclusion}

We have extended PH-based losses to the task of multi-class, CNN-based image segmentation.
Leveraging an enriched topological prior, including high dimensional and multi-class features, this approach improved segmentation topology in a post-processing framework.
Future work will seek to understand its limits with respect to the fidelity of CNN-based segmentation; its application within weakly supervised learning; and its extension to more challenging targets.


\bibliographystyle{splncs04}
\bibliography{references}

\end{document}